# Performance Evaluation of Java File Security System (JFSS)

**Brijender Kahanwal[1*], Tejinder Pal Singh[2] and R. K. Tuteja[3]**

[1]*Department of Computer Sc. & Engg., Shri Jagdishprasad Jhabarmal Tibrewala University, Jhunjhunu, Raj. (INDIA)*
[2]*Department of Physics, Shri Jagdishprasad Jhabarmal Tiberwala University, Jhunjhunu, Raj. (INDIA)*
[3]*N. C. Institute of Computer Science, Israna, Panipat, Haryana (INDIA)*

____________________________________________________________________________________________

**ABSTRACT**

*Security is a critical issue of the modern file and storage systems, it is imperative to protect the stored data from unauthorized access. We have developed a file security system named as Java File Security System (JFSS) [1] that guarantee the security to files on the demand of all users. It has been developed on Java platform. Java has been used as programming language in order to provide portability, but it enforces some performance limitations. It is developed in FUSE (File System in User space) [3]. Many efforts have been done over the years for developing file systems in user space (FUSE). All have their own merits and demerits. In this paper we have evaluated the performance of Java File Security System (JFSS). Over and over again, the increased security comes at the expense of user convenience, performance or compatibility with other systems. JFSS system performance evaluations show that encryption overheads are modest as compared to security.*

**Keyboard:** JFSS, Performance, Evaluation, File Security, File System.

____________________________________________________________________________________________

## INTRODUCTION

The operating system provides much fundamental functionality like storage management, memory management, process management, and the user interface. They also provide many security functionalities, but it always lacks the mechanism for the file security. When our Java File Security System (JFSS) is integrated once with the operating system, it enhances the file security on demand of the user. The file system is the primary focus of access control in an operating system. The flat file systems make poor secure file systems because there is no way to hide the existence of a file from a user. This approach is very convenient, and user friendly. It is developed in the user space and on the Java technology. The technology is well known for high portability, high CPU utilization by its multithreading feature, rich Application Programming Interface (API), and huge developer community.

For the sensitive data, it is important to have file level cryptographical access control. There are three types of cryptography in the file cryptographic systems, at file level, at file system level and at partition level. Every design has its own strengths and weaknesses when they are implemented. In the file level cryptography each file is encrypted or decrypted on the demand of the user. One example is AxCrypt [6] for Windows operating system. In the file system level cryptography, it provides security to the complete file system. Such type of systems encrypts all data that is going to the file system. All files are securely stored on the disk. This methodology is not independent of the underlying file system. These are very difficult to port. PGP [7] Whole Disk Encryption is the example for such type of systems. The third type of methodology comes in between the above two that is partition level cryptography. It has the special partition for the secure data or we can say that is the defined mount point where the secure files are to be kept. User will put the sensitive files in the partition on the demand. TrueCrypt [13] is the example of this type of cryptography.

Our system is JFSS that has some properties of the file level cryptography and which is implemented in the FUSE. It encrypts or decrypts the data files on the demand of the user and it can be mounted at any place on the disk. It also maintains the encryption key for encrypting or decrypting the file and that key is stored on the smart cards by the users. The encrypted file and the key are stored in the concern of the security of the data separately.

The rest of this article is organized as follows. In section 2 we discuss the reasons behind the development of the JFSS under the heading motivation. In section 3 the related work is described for the file systems. In section 4, we have evaluated the JFSS performance. In the section 5, we conclude and describe the future scope.

**Motivation:**
The file system is a major component of the operating system. It is a complex piece of software with layers below and above it, all affecting the performance of the system. Developing in-kernel file systems is a challenging task, because of many reasons [5]. These are as follows:

i)   The kernel code lacks memory protection,
ii)  It requires great attention to use of synchronization primitives,
iii) These can be written in C language only,
iv)  Debugging is a tedious task,
v)   Errors in the developed file systems can require rebooting the system,
vi)  Porting of the kernel file systems requires significant changes in the design and implementation, and
vii) These can be mounted only with super-user privileges.

But developing of file systems in user space eliminates all the above issues. At the same time, the research in the area of storage and file systems increasing involves the addition of rich functionalities over the underlying, as opposed to designing the low-level file systems directly. On the other end, by developing in user space, the programmer has a wide range of programming languages, third-party tools and libraries. The file system should be highly portable to other operating systems. The kernel remains smaller and more reliable. Because of these, we have developed the Java File Security System (JFSS). As the name describes, Java programming language has been selected for the development which is well known for the feature of high portability. There is also a potential disadvantage of user space file systems that they degrade performance of the system as compared to the kernel level implementation. There are additional context switches and memory copies overhead.

With the help of FUSE systems, we can implement the fully functional file system in user space program. There are so many features of FUSE development as simple library API, simple installation, secure implementation, user space - kernel interface is very efficient, usable by non privileged users, and has proven very stable over time etc. It is a loadable kernel module that helps in implementing the file systems in user space. Because of these we have make the choice of it. These are simple one to develop, but the encryption incurs more overhead than the kernel-space encryption. The figure 2.1 shows the FUSE module interactions in user space. Here JVM stands for Java Virtual Machine.

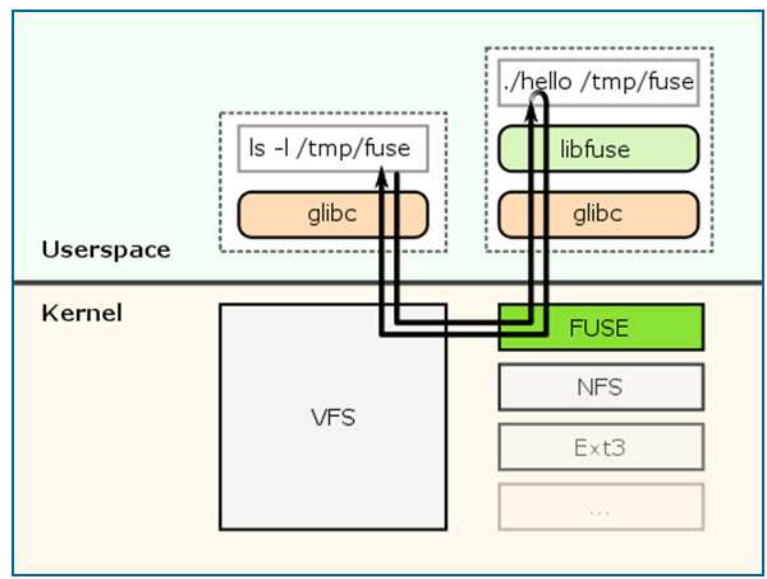

**Related Work:**
The idea of securing the stored data on disk with the help of encryption algorithms is well established. Many cryptographic file system projects have been implemented. Matt Blaze's CFS [2] is a most popular, portable user-level file system. It can encrypt any local directory on a system and that has different mount point. Users make choice of the encryption algorithms and a key to use. File data and metadata are encrypted. Its performance is limited by the number of context switches.

TCFS [4] is implemented as a kernel-mode cryptographic file system. It works transparently with the underlying file system. It uses less number of ciphers as compared to the CFS. All files are encrypted with the same cipher algorithm. It has two drawbacks first one is key is generated with the help of login passwords and second one is they are stored on the specific location in the system. All files of one user are encrypted with a single key. EncFS [11] is a encrypted file system that is implemented in the FUSE environment. It supports two ciphers namely AES and Blowfish. NCryptfs [12] is a stackable file system on the underlying file system. It is less transparent because it requires password for the accessing of the file. There is also much work done on the performances of the file systems [5].

**JFSS Performance:**

Here we evaluate JFSS from a performance perspective. There are three important factors for evaluating security systems: security, performance, and ease-of-use. Here we are concerned about the performance of the system. For our analysis we have created a simple benchmark to calculate the encryption time for the files. Security always takes higher costs in terms of space and time for any system. For the newly designed software or hardware, everyone is interested in its performance and features. That has significant impact on its popularity among the users.

Our fundamental aim in testing the performance of Java File Security System (JFSS) is basically to ensure that the security benefits do not come at too high cost. We carried out our experiments on Intel® Core™ 2 Duo CPU T6500 @ 2.10 GHz, 1.75 GB RAM, Windows XP with 120 GB hard drive machine.

We have clear goals for performance evaluation. How much space is acquired by the encrypted file and how much time is taken in the encryption process? What is the encryption overhead in terms of space? What are the relations between file size and execution time? What is the relation between file type and difference in the two sizes (original file size and encrypted file size)? These are the questions which are given answers after over evaluation of the Java File Security System (JFSS).

**Table 4.1 JFSS encryption overhead in terms of bytes and seconds for all types of files**

| Sr. No. | File Type | Size of Original File (Bytes) | File Size by JFSS (Bytes) | Encryption Overhead (Bytes) | Execution Time (Seconds) | Encryption Key Size (Bytes) |
|---|---|---|---|---|---|---|
| 1 | Text | 75 | 80 | 5 | 0.724 | 141 |
| 2 | Image | 5024 | 5040 | 16 | 1.234 | 141 |
| 3 | Excel | 8746 | 8752 | 6 | 1 | 141 |
| 4 | Bitmap | 20032 | 20048 | 16 | 0.967 | 141 |
| 5 | Document | 22016 | 22032 | 16 | 1.023 | 141 |
| 6 | Power Point Presentation | 27553 | 27568 | 15 | 1.132 | 141 |
| 7 | Executable | 43040 | 43056 | 16 | 1 | 141 |
| 8 | PDF | 905446 | 905456 | 10 | 2.933 | 141 |
| 9 | Audio | 9180972 | 9180976 | 4 | 67 | 141 |
| 10 | Video | 26246026 | 26246032 | 6 | 174 | 141 |

In this section, we show the table 4.1 which has different types of files details with varying sizes to encrypt them. The results are shown with the help of the table. It shows the details about the actual sizes of the files and after encryption sizes for the same files. It shows the difference or overhead of encryption in terms of storage space. It shows the execution time in seconds for the encryption process of particular file.

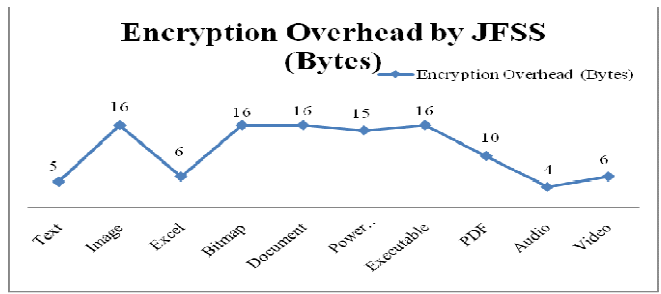

**Figure 4.1 JFSS encryption overhead in terms of bytes for occupying the strorage space.**

The figure 4.1 is drawn with the help of variables, file type and encryption overhead (along x-axis is file type and along y-axis is file size in bytes). It shows that the encryption overhead in terms of the difference between the two file types namely the original (normal size) file size and the size of the encrypted file. Both files are same one. The difference is maximum 16 bytes. It is not dependent of file sizes. It little much depends upon the type of file like text (5 Bytes), image (16 Bytes, excel (6 Bytes), and so on.

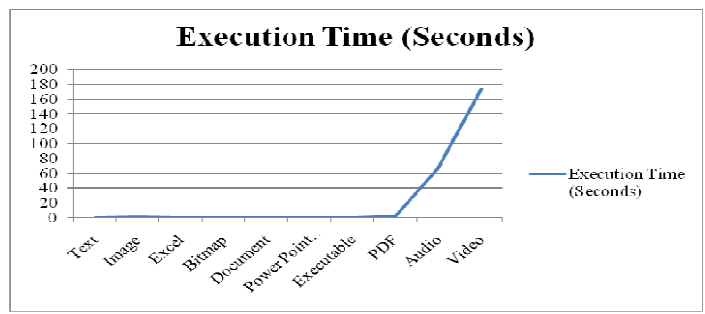

**Figure 4.2 The execution time (seconds) taken by the JFSS for encryption**

The figure 4.2 shows the execution time that is measured by the ad hoc microbenchmark program. In the graph, along the x-axis the file sizes are increasing and along the y-axis time is shown in seconds. As the file size is increasing, the encryption process's execution time is also increasing. The execution time variable is dependent on the file sizes.

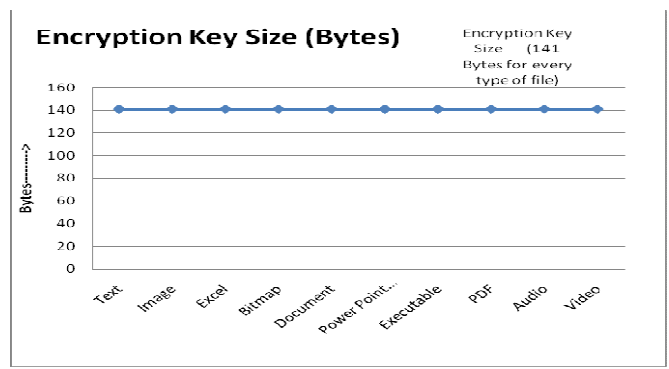

**Figure 4.3 The generated key sizes for every type of file in bytes by JFSS.**

In the figure 4.3, along the x-axis file types are taken and along the y-axis bytes are taken. The encryption key generated by the JFSS system is of size 141 Bytes. It is the standard size for the encryption key. It is not affected by the file sizes or by the file types. It is totally independent variable.

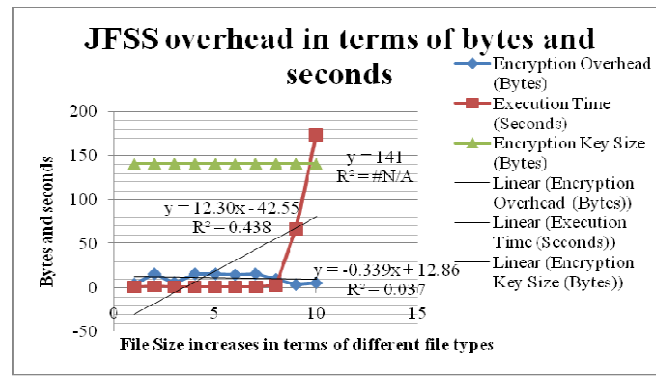

**Figure 4.4 Overall overhead of JFSS**

In the figure 4.4, the overall encryption overhead is described with the help of curve fitting. With the first degree polynomial equation line (y = ax + b), the relations are shown. For the encryption key size and the file types or sizes, there is no relation between these two. The encryption key size is an independent variable. For the variables file size difference and the file type the relation variable $R^2$ that is only 0.037 which is very week relation, we can say that is 16 bytes at most not more than that for any type of the file. The third relational value $R^2$ is 0.438, for the variables encryption execution time and file sizes. This shows that the two variables are dependent or related with each other. And this relation is also called the week one because that is in the range 0-0.5 range.

We conclude that the security overhead of the JFSS is very less on behalf of the security costs which are very high. The total memory space overhead is 157 bytes in total. It includes the file sizes difference that is max. 16 bytes and the encryption key size that is 141 bytes. It has the execution time overhead. The concepts say that we have to pay for the security. The system is very convenient to the users. The system is highly portable one.

**CONCLUSION**

We have shown the performance of our system JFSS for the file security. There is always a penalty of security that is the execution time taken for the encryption process and at the maximum 157 bytes (141Bytes for the encryption key and 16 Bytes of maximum size difference of normal file and an encrypted file) extra space acquired by the JFSS generated file( encrypted one). These are not considered against the security of the user data.

We will do the research work on the file system evaluation techniques. Basically the stress will be on trace file systems for the evaluation of the file and storage systems to work with the real workloads. In the future, the work should be done on the energy efficient file systems. They are the future technologies for the portable computer.

# REFERENCES

[1] Brijender Kahanwal, T. P. Singh, and R. K. Tuteja. *International Journal of Computer Science & Technology*, Vol. 2, Issue 3, pages 25-29, September **2011**.
[2] M. Blaze, "A Cryptographic File System for UNIX", in ACM Conference on Computer and Communications Security, pages 9-16, **1993**.
[3] FUSE-File System in User Space, http://fuse.sourceforge.net/.
[4] G. Cattaneo and G. Persiano. "Design and Implementation of a Transparent Cryptographic File System for UNIX", In Proceedings of the Annual USENIX Technical Conference, REENIX Track, pages 245-252, June **2001**.
[5] Aditya Rajgarhia and Ashish Gehani, "Performance and Extension of User Space File Systems",SAC'10 Proceedings of the 2010 ACM Symposium on Applied Computing, **2010**.
[6] AxCrypt. **2007**, Axantum Software AB. October 31 **2007**.
http://www.axantum.com/AxCrypt/.
[7] PGP Whole Disk Encryption. **2007**. PGP Corporation. October 31, **2007**.
http://www.pgp.com/products/wholediskencryption/.
[8] M. K. McKusick, W. N. Joy, S. J. Leffler, and R. S. Fabry. A fast file system for UNIX. ACM Transactions on Computer Systems, 2(3):181–197, August **1984**.
[9] M. Rosenblum and J. K. Ousterhout. The design and implementation of a log-structured file system. In Proceedings of 13th ACM Symposium on Operating Systems Principles, pages 1–15, Asilomar Conference Center, Pacific Grove, CA, October **1991**. Association for Computing Machinery SIGOPS.
[10] Andy Konwinski, John Bent, James Nunez, and Meghan Quist, "Towards an I/O Tracing Framework Taxonomy", In the Proceedings of Supercomputing'07 ACM Conference, November 10-16, **2007**.
[11] EncFS Encrypted File System. [Online]. http://www.arg0.net/encfs(**2010**/2).
[12] C. P. Wright, M. Martino, and E. Zadok, " NCryptfs: A Secure and Convenient Cryptographic File System", Proceedings of USENIX **2003** Annual Technical Conference, pages 197-210, **2003**.
[13] TrueCrypt Foundation. http://www.truecrypt.org, October 28, **2007**.
[14] Lavanya P and M Rajashekhara Babu, *Advances in Applied Science Research*, **2011**, 2 (3): 567-573.